\documentclass[aps,nofootinbib,showpacs,preprint]{revtex4}
\usepackage[CJKbookmarks, pdftex, bookmarksnumbered, bookmarksopen, colorlinks, citecolor=blue, linkcolor=blue]{hyperref}
\usepackage[english]{babel}
\usepackage{amstext,amsmath,amssymb,amsfonts,bbm}
\usepackage[latin1]{inputenc}
\usepackage{graphicx}
\usepackage{epstopdf}
\usepackage{amsthm}
\usepackage{tocvsec2}
\usepackage{enumerate}
\usepackage{subfigure}
\usepackage{color}
\usepackage{tikz}
\usepackage[squaren,Gray]{SIunits}

\newcommand{\be}{\nopagebreak[3]\begin{equation}}
\newcommand{\ee}{\end{equation}}

\newcommand{\bee}{\nopagebreak[3]\begin{equation*}}
\newcommand{\eee}{\end{equation*}}

\newcommand{\ba}{\nopagebreak[3]\begin{eqnarray}}
\newcommand{\ea}{\end{eqnarray}}
\DeclareFontFamily{U}{rsfs}{}         
\DeclareFontShape{U}{rsfs}{m}{n}{<5> rsfs5 <6><7> rsfs7          %
  <8><9><10><10.95><12><14.4><17.28><20.74><24.88> rsfs10}{}     %
\DeclareMathAlphabet{\mathfs}{U}{rsfs}{m}{n}                     %
\newcommand{\mfs}[1]{\mathfs {#1}}                               %

\newcommand{\sO}{{\mfs O}}

\begin{document}

\title{A microscopic model for an emergent cosmological constant}

\author{Alejandro Perez}
\affiliation{{Aix Marseille Univ, Universit\'e de Toulon, CNRS, CPT, Marseille, France\\
{ perez@cpt.univ-mrs.fr}}}
\author{Daniel Sudarsky}
\affiliation{Instituto de Ciencias Nucleares, Universidad Nacional Aut\'onoma de M\'exico, M\'exico D.F. 04510, M\'exico\\
{ sudarsky@nucleares.unam.mx} }

\author{James D. Bjorken}
\affiliation{SLAC National Accelerator Laboratory,
Stanford University, Stanford, California 94309, USA\\
bjbjorken@gmail.com }
\date{\today}

\begin{abstract}

The value of the cosmological constant is explained in terms of a noisy diffusion of energy from the low energy particle physics degrees of freedom to the fundamental Planckian granularity which  is  expected from general arguments in quantum gravity. The quantitative success of our phenomenological model is encouraging and provides possibly useful insights about physics at the scale of quantum gravity. 
 \\
 \begin{center}
{\em \bf  Essay written for the Gravity Research Foundation 2018 Awards for Essays on Gravitation.}
 \end{center}
 \end{abstract}

\maketitle

Naive vacuum energy estimates of the value of  the cosmological constant produces results that are 120  orders of magnitude larger than those extracted  from  observations \cite{Adam:2015rua}. Alternative  considerations  based on protective symmetries set their value to zero.   The usual  analysis largely  relies   on standard  notions of spacetime and fields which are smooth to all scales; however,  various arguments  suggest that the physics of the continuum should only emerge from an underlying discrete reality at the fundamental scale. We will remain agnostic  about the exact nature of  the underlying fundamental physics, yet we will take seriously the central idea that nature is discrete at the Planck scale. We shall argue that this assumption opens the door for a fresh new look into the cosmological constant problem.

The molecular structure of matter has important macroscopic consequences. 
The  most  ubiquitous  being  friction: the generic tendency of energy to leak from macroscopic scales to the underlying microscopic chaos.  Similarly, if the continuum is emergent, energy should `diffuse' from the large-scale  degrees of freedom (represented by matter fields in spacetime) down to the fundamental granular structure.  For reasons stated below,  the diffusion we postulate is tied to the presence of a non-trivial curvature. This makes it virtually undetectable in local laboratory searches involving essentially flat spacetime regions  
  \footnote{The simplistic view where Planckian discreteness is tied to a globally  defined  preferred  frame seems very tightly constrained \cite{Mattingly:2005re, Collins:2004bp}.
The idea, inspiring in part the present model, that the granularity should rather be associated  
 to a frame locally determined by geometrical features (e.g. curvature) and/or  the matter distribution and some of its  phenomenological implications susceptible  to laboratory testing were considered in \cite{Corichi:2005fw, Bonder:2007bj, Bonder:2008et,  Terrano:2011zz,  Aguilar:2012ju}.}. 
The  exception however is provided by the physics of the very early universe, where curvature and matter densities become large.  

For an individual particle,  such  diffusion effects are most reasonably encoded in a deviation from geodesic motion.
The discreteness that sources friction is associated with the Planck scale $\ell_p=m_p^{-1}$ which, in a relational spirit, must be identified within the rest frame of the particle which only exists if the excitation is massive. The presence of massive degrees of freedom (and the associated breaking of scale invariance) is signaled by a non-vanishing value of the trace of the energy momentum tensor ${\mathbf T}=g^{\mu\nu} {\mathbf T}_{\mu\nu}$ which, via Einstein's equations \footnote{In our model Einstein's equations suffer only small corrections.}, 
can be related to the scalar curvature ${\mathbf R}=-8\pi G {\mathbf T}$.

Therefore, it is natural to postulate that such a `friction' force must be proportional to ${\mathbf R}$. In addition, the force should depend on the mass $m$, the 4-velocity $u^{\mu}$, the spin $s^{\mu}$ of the classical particle (the only intrinsic features defining a particle), and a time-like unit  vector  $\xi^{\mu}$ specifying the local frame  defined  by the matter that curves spacetime.  Dimensional analysis gives an essentially  unique expression (see \cite{Perez:2017krv} for more details)  which  is   compatible with the above requirements \footnote{Higher curvature corrections could be added, but these are highly suppressed by the Planck scale and are thus negligible for the central point of this essay.} 
\ba\label{modimodi}
u^{\mu}\nabla_{\mu}  u^{\nu}&=&-\alpha\, m\, {\rm sign}(s\cdot \xi) \frac{ {\mathbf R}}{m^2_p}\, {s^{\nu}}   
\ea
where $\alpha> 0$ is a dimensionless coupling  \footnote{It is important to point out that the violations of the equivalence principle and Lorentz invariance implied by \eqref{modimodi} can be checked to avoid conflict with present observational bounds by many orders of magnitude \cite{Kostelecky:2008ts}. A simple indication comes   from  comparison  of   the values of curvature at the  EW transition in  cosmology to  that associated  with, say, the gravitational effect  of a  piece of lead, which gives $\frac{{\mathbf R} _{lead}} {{\mathbf R}_{EW}}  \sim 10^{-24}$.}.

 In cosmology $\xi=\partial_t$
is the time-arrow of the  co-moving cosmic  fluid.
The factor ${\rm sign}(s\cdot \xi)$  makes the force genuinely friction-like.
  This is apparent when one computes the behaviour of the mechanical energy of the particle  $E\equiv -m u^{\nu}\xi_{\nu}$ (in the frame defined  by $\xi^{\mu}$) which yields
\be\label{conserv} \dot E\equiv -m u^{\mu}\nabla_{\mu}(u^{\nu}\xi_{\nu})=-\alpha  
\frac{m^2 }{m^2_p} |(s\cdot \xi)| {\mathbf R}-m u^{\mu}u^{\nu}\nabla_{(\mu}\xi_{\nu)}.\ee
The last term  in \eqref{conserv} encodes the  standard  change of $E$  associated to the  non-Killing  character of  $\xi^{\mu}$.
The first term on the right  encodes the friction that damps out any motion with respect to $\xi^\mu$.  Energy is lost into the fundamental granularity until the particle is at rest with the cosmological fluid, i.e., $u^{\mu}=\xi^{\mu}$, and thus $\dot E=0$.

Fermions are fundamentally quantum objects which arguably interact directly with the physics at the Planck scale. The presence of spin on the r.h.s. of (\ref{modimodi}) is of course consistent with this view. Another important peculiarity of fermions is captured by the spin-statistics theorem (Pauli's exclusion principle). 
Fermions are sources of torsion which can be viewed as local defects in the Riemannian geometry 
\footnote{\label{cinco}
In this respect, notice that  the characterization of WKB-trajectories of the Dirac theory on a pseudo-Riemannian geometry \cite{Audretsch:1981xn} which, to lowest order in $\hbar$, is given by
\be\label{wkb}
u^\nu \nabla_\nu (m u_\mu) =-\frac{1}{2} \tilde {\mathbf R}_{\mu\nu \rho  \sigma} u^\nu \langle S^{\rho\sigma}\rangle +\sO(\hbar^2).
\ee
The previous is aquivalent to (\ref{modimodi}) if we introduce an effective $\tilde {\mathbf R}_{\mu\nu \rho  \sigma}\propto  {m^2}/{m_p^2}\,  {\rm sign}(s\cdot\xi) {\mathbf R} \epsilon_{\mu\nu \rho  \sigma}$ which encode a pure torsion structure as $\tilde {\mathbf R}_{[\mu\nu \rho]  \sigma}\not=0$ (from the first Bianchi identities). Note also the similarity between (\ref{modimodi}) and the Mathisson-Papapetrou-Dixon equations  \cite{Papapetrou:1951pa}. 
}
.  This implies that Fermions require the use of tetrads and connections $ ( e, \omega )$  when considering their coupling to gravity (e.g. the Einstein-Cartan formalism or any of its generalizations). The torsion part of $\omega$ can be integrated-out producing effective Fermi four-fermion-interactions with a coupling constant $\ell_p^2=m_p^{-2}$ \cite{Perez:2005pm, Freidel:2005sn}. As for gravity itself, the emergence of such non-renormalizable interactions implies that a consistent gravitational coupling of fermions must necessarily have an effect at the fundamental scale.  This has led to appealing pure-fermion proposals for the unification of interactions \cite{1959ZNatA..14..441D, Nambu:1961fr, Bjorken:1963vg}. In non-perturbative approaches to quantum gravity (e.g. loop quantum gravity \cite{Rovelli:2004tv})  Fermions produce Planck scale defects in the quantum geometry.   We have become accustomed to treat  them through the use of Grassmann variables in  path integrals, but, in our  view,  Fermions remain  enigmatic objects whose intrinsic nature is tightly related to the fundamental structure of spacetime. 
All this naturally suggests that Fermions interact directly with the  Planckian granular structure from which classical spacetime is expected to emerge. Although Eqn. (\ref{modimodi}) is expressed in quasi-classical terms,  which, for fermions, is strictly speaking problematic (see however footnote \ref{cinco}), we nevertheless find it significant that it encodes our expectations in such a natural way.
%

The non-conservative force in (\ref{modimodi}) leads to  a specific  form of  violation of  the energy momentum tensor describing  a fluid composed of  such particles. Yet this violation is inconsistent  with general relativity (GR). Fortunately,  there is a slight generalization of GR \cite{Einstein1919Spielen-Gravita}, called  {\it unimodular gravity} (UG),  where  this limitation is overcome \cite{Josset:2016vrq}. In UG  Einstein's equations  are replaced by the trace free equations
\begin{equation}\label{TraceFreeEinsteinEquation}
	{\mathbf R}_{\mu\nu} - \frac{1}{4} {\mathbf R}\, g_{\mu\nu} = {8 \pi G} \left( {\mathbf T}_{\mu\nu} - \frac{1}{4} {\mathbf T} g_{\mu\nu} \right).
\end{equation}
Defining $J_\mu\equiv (8\pi G) \nabla^\nu {\mathbf T}_{\nu\mu}$, assuming the unimodular integrability $dJ=0$ \cite{Josset:2016vrq}, and using Bianchi identities, one obtains
\begin{equation}\label{TraceFreeEinsteinEquation2}
{\mathbf R}_{\mu\nu} - \frac{1}{2} {\mathbf R} \, g_{\mu\nu} +\underbrace{\left[\Lambda_{ *} + \int_{\ell} J\right] }_{\rm Dark \ Energy\ \Lambda}g_{\mu\nu}= {8 \pi G} {\mathbf T}_{\mu\nu} ,
\end{equation}
where $\Lambda_{*}$ is an integration  constant.  Note that the current $J$  sources  a term in effective Einstein's equations that behave as dark energy \footnote{As noted  in \cite{Weinberg:1988cp, Ellis:2010uc}, `vacuum energy' needs not gravitate  in the  context  of UG,  aleviating  the tension  between quantum field theory and cosmology \cite{Ellis:2013eqs}. }.

\begin{figure}[h]
\centerline{\hspace{0.5cm} \(
\begin{array}{c}
\includegraphics[height=4.5cm]{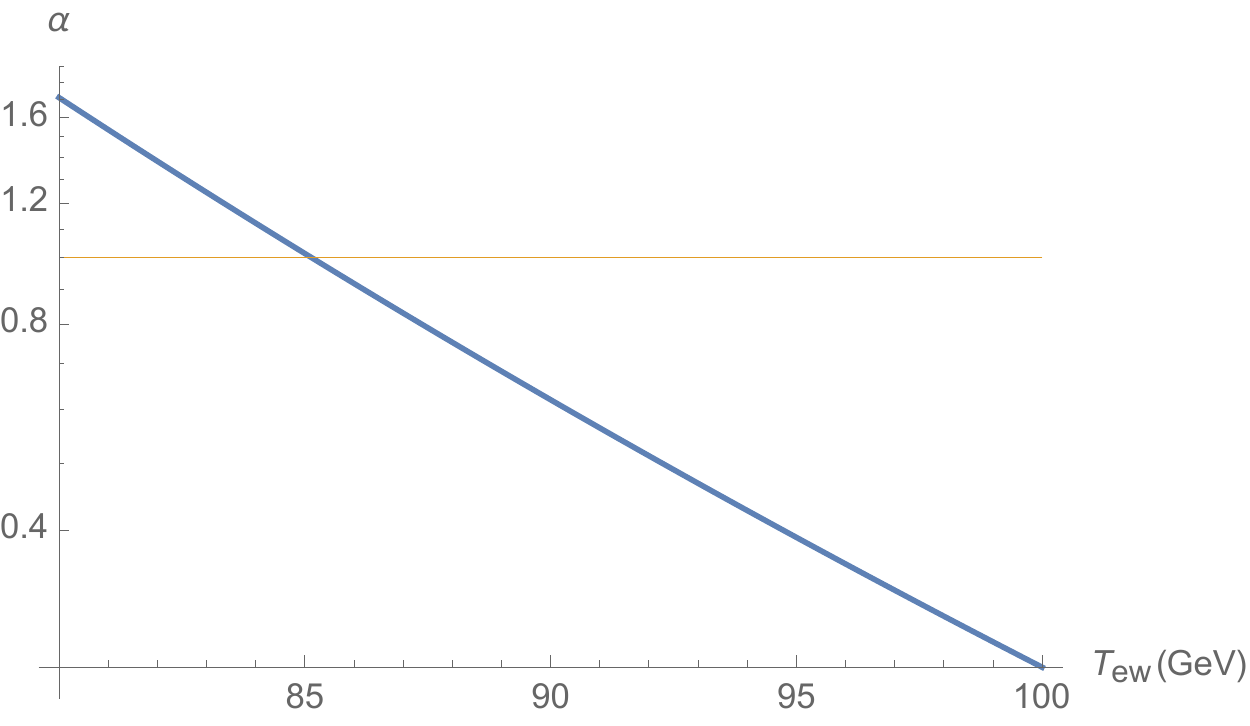} 
\end{array}\ \ \ \ \ \ \  \begin{array}{c}
\includegraphics[height=5cm]{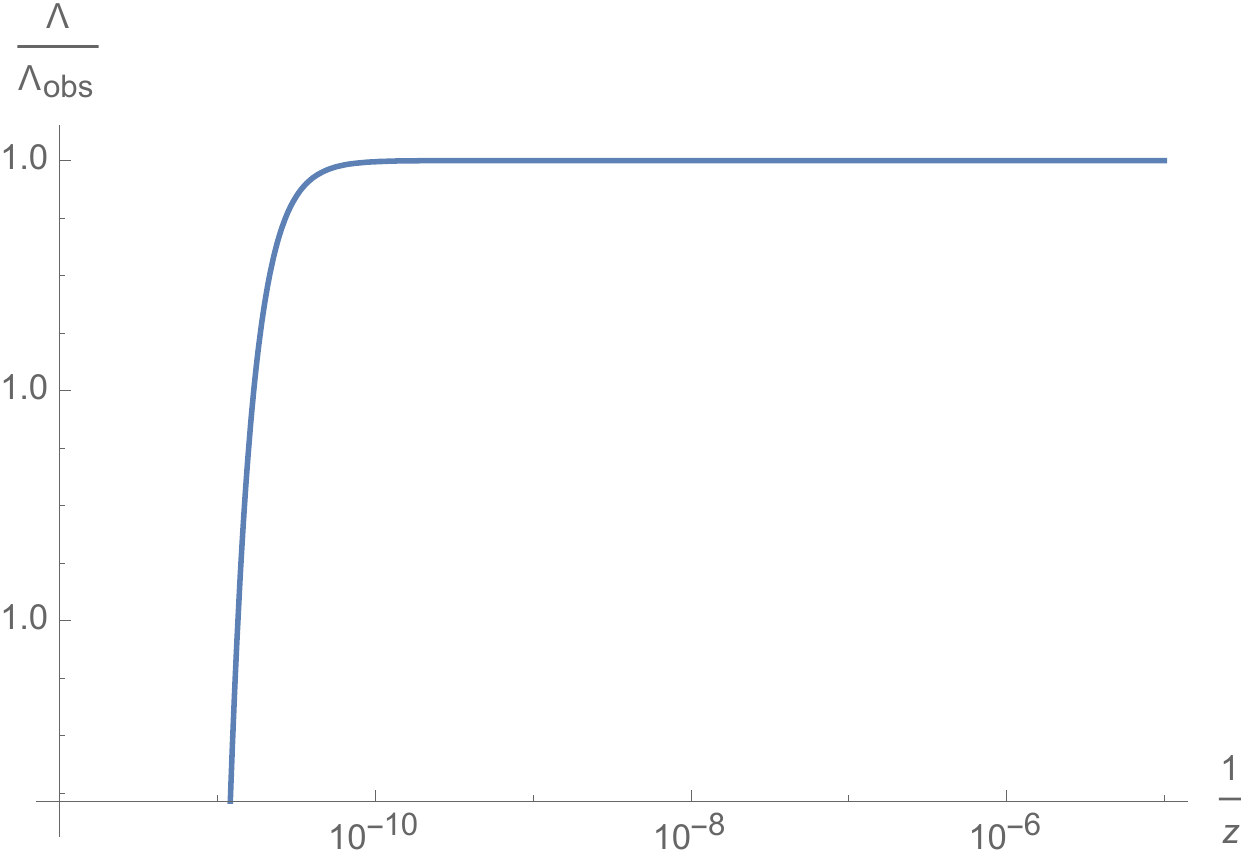} 
\end{array}\) } \caption{{\bf Left:} The value of the phenomenological parameter $\alpha$ that fits the observed value of $\Lambda_{obs}$ as a function of the EW transition scale $T_{\rm ew}$ in $\rm GeV$. We see that for $T_{\rm ew}\approx 100GeV$ $\alpha \approx 1$. {\bf Right:} The time dependence of $\Lambda$ expressed in terms of the inverse redshift factor $1/z$. 
}
\label{figu}
\end{figure}

Let us  focus  on the  dynamics of the early universe when its macroscopic geometry is well approximated by the flat  FLRW metric 
\be
ds^2=-dt^2+a(t)^2 d\vec x^2,
\ee
and where the local frame $\xi=\partial_t$ is  identified  with comoving observers.
The cosmological fluid is defined by an ensemble of quasi-classical particles characterized by a distribution in phase space satisfying a relativistic Boltzman equation with the external force coming from (\ref{modimodi}) \cite{rezzolla2013relativistic}. With this one can compute the   exact form of $J_\nu \equiv {(8\pi G)\, \nabla^\mu {\mathbf T}_{\mu\nu}}$  in  thermal  equilibrium at temperature $T$.  Considering  $T\gg m$  (the relevant regime for what follows) one obtains: 
\be
J_\nu\equiv(8\pi G)\, \nabla^\mu {\mathbf T}_{\mu\nu}=-{2 \pi \alpha \hbar} \frac{T  }{m_p^2}  {\mathbf R}^2 \xi_ \nu.
\ee
 As only massive  particles  with spin are  subjected  to the 
frictional force  \eqref{modimodi},  the diffusion mechanism in cosmology starts when such particles first appeared. According to the standard model---whose validity is assumed from the end of inflation---this corresponds to the electro-weak (EW) transition time.
 We further assume that a protective symmetry enforces  $\Lambda_*=0$ (see for instance \cite{Hawking:1984hk,Coleman:1988tj}).    
From (\ref{TraceFreeEinsteinEquation2}) we get
\be\label{cosmo}
\Lambda =\frac{2\pi \alpha \hbar}{m_p^2}\left(\int_{t_0}^{\rm t} T(t) {\mathbf R}(t)^2  dt\right) 
\ee
The integral in (\ref{cosmo}) can be performed using the standard model of cosmology.  The results  are shown in Figure \ref{figu}: the dark energy component in \eqref{cosmo}  rapidly approaches a constant, and its value fits the observed value  for $\alpha\approx 1$ and $T_{ew}\approx 100\, {\rm GeV}$. A rough estimate of the calculation  can be expressed in terms of the scales involved. The result is dominated by the top quark; the most massive excitation at $T_{ew}$, thus
\be
\Lambda \approx \frac{\overline m_t^4 T_{ew}^3}{m^7_p} m_p^2\approx \underbrace{\left(\frac{
T_{ew}}{m_p}\right)^7}_{10^{-120}} m_p^2,
\ee where $\overline m_t$  is the mean top mass $m_t$ that goes from zero to $170 \, {\rm GeV}$ during the EW-transition\footnote{\label{fn} Massive gauge bosons do not change the order of magnitude estimate, as ${m_Z}/{m_t}\approx 1/2$ and ${g_{ZW^{\pm}}}/{g_{t\bar t}}=3/4$. In (\ref{cosmo}) this leads to a factor $(3/4)^2 (1/2)^4$ which is about only $3.5 \%$ of the top-quark contribution.}. We have taken $\overline m_t\approx T_{ew}$ in the second approximation. This indicates that the present accelerated expansion of the universe might   simply be the result  of  a noisy diffusion of energy into the underlying granularity of spacetime. 


%

Torsion might be the simplest and most direct characterization of the defects in the emergent spacetime geometry that is responsible for our diffusion effect. But it can be at best only a small part of a bigger, as yet unknown, story.
 
The situation may be likened to that present during the birth of the quantum theory. From the very beginning, `discretization' of phase space was strongly indicated (e.g. the Rayleigh-Jeans catastrophe). However, going from simple subdivision of phase space into cells {\em \'a la} Boltzmann to Dirac's ultimate formulation (`replace Poisson brackets by commutators') required three decades of a great deal of experimental and theoretical work. With regard to spacetime discretization, we should expect no less.

The above quantum-theory example underlines the importance of finding the descriptive language most appropriate for the problem at hand. In addition to the relevance of torsion, as expressed in the first order O(3,1) gauge-gravity formalism, there are other options available. The relevance of the vacuum topological structure is suggested by the O(4,1) extension of gauge gravity, as well as loop quantum gravity. Quantized areas and volumes are also suggested by loop quantum gravity. And a variety of theory extensions (e.g. M-theory) are suggested by string theory.

We are used to the expectation that discoveries of physics at microscopic scales often provide clues about physics at larger scales. But the reverse happens as well. Could it be that in this case the physics of the largest scales (cosmology) is providing us with clues about the physics at extremely small scales? We might be reassured of this possibility by considering the manner in which regularities found in the study of chemical reactions uncovered the first empirical insights for what would become the atomic theory, in contrast with the purely conceptual considerations that inspired old dear Democritus.



\begin{thebibliography}{10}

\bibitem{Adam:2015rua}
R.~Adam et~al.
\newblock {Planck 2015 results. I. Overview of products and scientific
  results}.
\newblock {\em Astron. Astrophys.}, 594:A1, 2016.

\bibitem{Mattingly:2005re}
David Mattingly.
\newblock {Modern tests of Lorentz invariance}.
\newblock {\em Living Rev. Rel.}, 8:5, 2005.

\bibitem{Collins:2004bp}
John Collins, Alejandro Perez, Daniel Sudarsky, Luis Urrutia, and Hector
  Vucetich.
\newblock {Lorentz invariance and quantum gravity: an additional fine-tuning
  problem?}
\newblock {\em Phys.Rev.Lett.}, 93:191301, 2004.

\bibitem{Corichi:2005fw}
Alejandro Corichi and Daniel Sudarsky.
\newblock {Towards a new approach to quantum gravity phenomenology}.
\newblock {\em Int.J.Mod.Phys.}, D14:1685--1698, 2005.

\bibitem{Bonder:2007bj}
Yuri Bonder and Daniel Sudarsky.
\newblock {Quantum gravity phenomenology without Lorentz invariance violation:
  A Detailed proposal}.
\newblock {\em Class. Quant. Grav.}, 25:105017, 2008.

\bibitem{Bonder:2008et}
Yuri Bonder and Daniel Sudarsky.
\newblock {Unambiguous Quantum Gravity Phenomenology Respecting Lorentz
  Symmetry}.
\newblock {\em Rept. Math. Phys.}, 64:169--184, 2009.

\bibitem{Terrano:2011zz}
W.~A. Terrano, B.~R. Heckel, and E.~G. Adelberger.
\newblock {Search for a proposed signature of Lorentz-invariant spacetime
  granularity}.
\newblock {\em Class. Quant. Grav.}, 28:145011, 2011.

\bibitem{Aguilar:2012ju}
Pedro Aguilar, Daniel Sudarsky, and Yuri Bonder.
\newblock {Experimental search for a Lorentz invariant spacetime granularity:
  Possibilities and bounds}.
\newblock {\em Phys. Rev.}, D87(6):064007, 2013.

\bibitem{Perez:2017krv} 
  A.~Perez and D.~Sudarsky,
  ``Dark energy from quantum gravity discreteness,''
  arXiv:1711.05183 [gr-qc].


\bibitem{Kostelecky:2008ts}
V.~Alan Kostelecky and Neil Russell.
\newblock {Data Tables for Lorentz and CPT Violation}.
\newblock {\em Rev. Mod. Phys.}, 83:11--31, 2011.

\bibitem{Audretsch:1981xn}
J.~Audretsch.
\newblock {Dirac Electron in Space-times With Torsion: Spinor Propagation, Spin
  Precession, and Nongeodesic Orbits}.
\newblock {\em Phys. Rev.}, D24:1470--1477, 1981.

\bibitem{Papapetrou:1951pa}
Achille Papapetrou.
\newblock {Spinning test particles in general relativity. 1.}
\newblock {\em Proc. Roy. Soc. Lond.}, A209:248--258, 1951.

\bibitem{Perez:2005pm}
Alejandro Perez and Carlo Rovelli.
\newblock {Physical effects of the Immirzi parameter}.
\newblock {\em Phys. Rev.}, D73:044013, 2006.

\bibitem{Freidel:2005sn}
Laurent Freidel, Djordje Minic, and Tatsu Takeuchi.
\newblock {Quantum gravity, torsion, parity violation and all that}.
\newblock {\em Phys. Rev.}, D72:104002, 2005.

\bibitem{1959ZNatA..14..441D}
H.P. {D{\"u}rr}, W.~{Heisenberg}, H.~{Mitter}, S.~{Schlieder}, and
  K.~{Yamazaki}.
\newblock {Zur Theorie der Elementarteilchen}.
\newblock {\em Zeitschrift Naturforschung Teil A}, 14:441--485, June 1959.

\bibitem{Nambu:1961fr}
Yoichiro Nambu and G.~Jona-Lasinio.
\newblock {Dynamical model of elementary particles based on an analogy with
  superconductivity. II}.
\newblock {\em Phys. Rev.}, 124:246--254, 1961.

\bibitem{Bjorken:1963vg}
J.~D. Bjorken.
\newblock {A Dynamical origin for the electromagnetic field}.
\newblock {\em Annals Phys.}, 24:174--187, 1963.

\bibitem{Rovelli:2004tv}
Carlo Rovelli.
\newblock {\em {Quantum gravity}}.
\newblock Cambridge Monographs on Mathematical Physics. Univ. Pr., Cambridge,
  UK, 2004.

\bibitem{Einstein1919Spielen-Gravita}
A.~Einstein.
\newblock {Spielen Gravitationsfelder im Aufbau der materiellen
  Elementarteilchen eine wesentliche Rolle?}
\newblock {\em Sitzungsber. Preuss. Akad. Wiss. Berlin}, pages 349--356, 1919.

\bibitem{Josset:2016vrq}
Thibaut Josset, Alejandro Perez, and Daniel Sudarsky.
\newblock {Dark energy as the weight of violating energy conservation}.
\newblock {\em Phys. Rev. Lett.}, 118(2):021102, 2017.

\bibitem{Weinberg:1988cp}
Steven Weinberg.
\newblock {The Cosmological Constant Problem}.
\newblock {\em Rev. Mod. Phys.}, 61:1--23, 1989.

\bibitem{Ellis:2010uc}
George F.~R. Ellis, Henk van Elst, Jeff Murugan, and Jean-Philippe Uzan.
\newblock {On the Trace-Free Einstein Equations as a Viable Alternative to
  General Relativity}.
\newblock {\em Class. Quant. Grav.}, 28:225007, 2011.

\bibitem{Ellis:2013eqs}
George F~R Ellis.
\newblock {The Trace-Free Einstein Equations and inflation}.
\newblock {\em Gen. Rel. Grav.}, 46:1619, 2014.

\bibitem{rezzolla2013relativistic}
L.~{Rezzolla} and O.~{Zanotti}.
\newblock {\em {Relativistic Hydrodynamics}}.
\newblock {Oxford University Press}, Oxford, UK, 2013.

\bibitem{Hawking:1984hk}
S.~W. Hawking.
\newblock {The Cosmological Constant Is Probably Zero}.
\newblock {\em Phys. Lett.}, 134B:403, 1984.

\bibitem{Coleman:1988tj}
Sidney~R. Coleman.
\newblock {Why There Is Nothing Rather Than Something: A Theory of the
  Cosmological Constant}.
\newblock {\em Nucl. Phys.}, B310:643--668, 1988.

\end{thebibliography}
 \end{document}